# A New Index of Human Capital to predict Economic Growth

Henry Laverde

henrylaverde@usantotomas.edu.co

Universidad Santo Tomás, Facultad de Economía

Bogota, Colombia

Juan C. Correa

Facultad de Psicología. Fundación Universitaria Konrad Lorenz

Bogotá, Colombia

Klaus Jaffe

Departamento de Biología de los Organismos. Laboratorio de Comportamiento

Universidad Simón Bolívar, Caracas, Venezuela.

## Clarifications and limitations of the proposed model

Human capital is an abstract concept that encompasses individual cognitive abilities and skills. To identify human capital, this document uses the position of it in the productive process. On the one hand, it is known that to reach a certain level, a number of inputs must be used, such as schooling and health. On the other hand, once a certain level of human capital is attained, a series of returns must be generated and reflected in productivity, capacity for innovation, etc. In this sense, human capital is situated in the middle of the productive process, such that it is neither an input nor a final product. However, because human capital is not observable, it can be approximated through related inputs or returns associated with individuals, with the hope that there is a high correlation between inputs or returns and human capital. Both approaches have been used in the literature, primarily with one-dimensional structures.

Independently of the approach used, the question arises as to why a composite measure if a single variable could be used to approximate human capital. To answer this question, average years of education are used as an example for the input approach and the two variables used in this document, EC and PP, are used as examples for the returns approach.

For one of these variables to be considered a reliable approximation of human capital, it should account for the largest percentage of the variance of this stock. Therefore, under the input approach, movements in average years of education should translate into movements in human capital in the same direction and with the largest possible magnitude. The greater the correlation between the movements of these two forces, the more certain it is that human capital can be captured with a single variable, regardless of the inability to observe human capital.

Regarding the average years of education variable, several studies have found that this variable does not satisfactorily meet this goal. The first argument against this variable is that, as the only element to approximate human capital, it does not correct for quality in assuming homogeneity among individuals and thus makes individuals perfect substitutes for one another. For the same reason, it makes it impossible to account for differences between countries' education systems across space and time. Finally, categorizing human capital according to educational level omits the human capital of individuals who lack formal education. Thus, average years of education by itself may not be a good approximation of human capital, particularly to the extent that these limitations gain significance and transform this variable into a mere reflection of credentialism, which does not translate into an increase in the skills and capabilities of individuals.

The problem with the productivity variable (EC) is that variations in productivity do not necessarily correspond exclusively to variations in human capital. Rather, productivity can also vary due to other factors (physical capital, labor, etc.). Without additional elements, this measure of human capital is imperfect and might even capture distant elements, which generates an upward bias if national productivity is driven significantly by other elements, such as physical capital.

In principle, variance in the innovation variable (PP) seems to be explained nearly exclusively by variance in human capital, but the use of innovation as a single proxy is also limited. The issue is that this variable captures the portion of human capital that is concentrated at the top of the distribution. Although the generation of new knowledge or technologies requires highly qualified individuals, this variable excludes human capital that does not necessarily translate into innovation but is important in other areas, such as productivity or the absorption of new technology. In this sense, the problem with PP is similar to that of average years of education; that is, it attributes low levels of human capital to countries with low levels of PP.

A more traditional variable used in the returns approach is international cognitive test scores. This measure collects international test scores in mathematics and natural sciences. The convenience of using academic results in the areas of math and science has been justified on the basis that these disciplines are closely related to innovation and productivity, from both theoretical and empirical perspectives (Hanushek and Kimko, 2000). However, this variable presents several problems that hinders its consistent use in empirical applications, especially longitudinal applications. Specifically, the number of developing countries in the sample is reduced and the number of periods in the study is very low. From a conceptual point of view, this variable also presents certain limitations that reduce its ability to approximate human capital by itself. First, the degree to which scores on academic tests in math and science capture variations in productivity or innovation is debatable. The skills and abilities of workers manifest in many respects and cannot be reduced to a small number of qualities, such as the ability to solve mathematical problems. Although this measure corrects for quality, it also, to some extent, assumes the homogeneity of national educational systems. However, certain educational systems might place greater importance on other types of skills, thereby reducing national test results. Furthermore, because technical and tertiary education are key to individuals' productivity and innovation, one could argue that good results in basic and intermediate education do not necessarily guarantee good results in the later stages of education, particularly in developing countries, where the dropout rates are very high.

In summary, the one-dimensional approach imperfectly measures human capital, and the errors will increase to the extent that the limitations mentioned above become more important. For example, the data show that countries are tending to homogenize in terms of the average years of education, which should translate into homogenization of performance; however, gaps in productivity, innovation and academic test scores remain significant.

One way to overcome the drawbacks of the one-dimensional approach is to combine these variables in a manner that extracts the components linked exclusively to human capital in a more comprehensive and systematic way while maintaining theoretical coherence and avoiding the creation of circularity problems. The challenge is to identify the parts of these variables that are attributable to human capital when the input and returns approaches are distinguished and merged. This is the precise objective of the model presented in this paper. To illustrate this point, let us suppose that human capital (*Hc*) can be expressed as

$$Hc = \lambda IV + u \qquad (1)$$

where *Hc* is explained by an input variable *IV* (such as, for example, average years of education) and by a vector of other factors $u$. For *IV* to be considered a good approximation of *Hc*, it should be expected that $u$ is very close to 0, that is, $\mathrm{E}(u) = 0$, and that $\lambda$ is as high as possible. On the contrary, if the factors contained in $u$ significantly affect $Hc$, the input variable loses validity as a single element to approximate $Hc$. In the case of average years of education, the greater the differences in educational systems between countries, the higher $u$ will be and the less valid the average years of education variable will be as a proxy.

Likewise, a returns variable ($RV$) can be expressed as:

$$RV = \delta Hc + \mu \qquad (2)$$

The higher the value of $\mu$ is, the lower the explanatory power of human capital for the outcome variable, which reduces the relevance of $RV$ as a proxy. For example, if a large portion of the variation of the productivity variable is the result of physical capital, $\mu$ is higher and the validity of $RV$ is decreased.

Now, it is possible to assume that we have two variables of returns,

$$VR_1 = \delta_1 Hc + \mu_1 \qquad (3)$$

$$VR_2 = \delta_2 Hc + \mu_2 \qquad (4)$$

These two variables share a common component, namely, human capital. Accordingly, if it is assumed that $\mathrm{E}(\mu_1 \mu_2) = 0$, the movements of each $RV_i$ will be in different directions as a result of $\mu_i$ and in the same direction as a result of $Hc$. If variables with these features are found, then $Hc$ can be identified by extracting the shared variance between the two variables. This would be possible provided that $\mathrm{E}(\mu_1 \mu_2) = 0$ and that the correlation between the $RV_i$ is high. Superficially, this is precisely what methodologies such as principal components analysis (PCA) or PLS-PM do; they utilize the correlation between variables that measure the same concept to identify an underlying latent variable. These methodologies construct a variable composed of the weighted sum of the original variables, maximizing the shared variance between them and striving to lose the least amount of information. However, PLS-PM goes further by allowing the construction of a system of equations that merges the input and returns approaches and clearly distinguishes the variables that belong to each of these approaches.

This approach has advantages over other methodologies because the HCI is estimated under the returns approach but incorporates feedback from variables of the input approach, which not only allows a construction of the concept from a perspective of equilibrium but also provides more theoretical consistency. A methodology such as ACP could also be used, but this methodology would be theoretically inconsistent if no distinction is made between variables of inputs or variables of returns; therefore, it would be difficult to establish with accuracy what is being calculated. In addition, ACP could generate a problem of circularity, thereby exacerbating the estimate bias.

Moreover, it is clear that as the number of manifest variables increases, the quality of the estimate improves. However, the number of variables included in the model is restricted by the availability and characteristics of the data. The problem with the model as defined by equation 1 to 4 relates to the block that encompasses returns on human capital; hence, the question arises whether EC and PP are sufficient to approximate human capital.

As discussed above, neither of these variables alone is sufficient to approximate human capital. By combining them and extracting the shared variance corresponding to the returns on human capital, a high degree of association between the manifest variables should result, as well as knowledge of what is being measured. Evaluation with the one-dimensional and discriminant test of the PLS-PM model ensures the first point. Regarding the second point, note that the variable PP contains only the top part of the distribution of human capital; specifically, individuals with high qualifications. Consequently, when the portion that corresponds to human capital using PP as reference is extracted from EC and vice versa, only a fragment of this portion would be extracted, which generates a bias in the estimates.

To avoid this possible bias, one or more manifest variables should capture the full spectrum of human capital. Other authors have included variables such as trade in technological equipment (the sum of imports and exports) to capture the ability of individuals to absorb and adapt to new technology (Messinis and Ahmed, 2013) and cognitive test scores (Hanushek and Woessmann, 2008) as complements, but given the limited sample size and time periods for these data, they will only be useful in this document for purposes of cross-section and sensitivity analysis.

# Tests to validate the results

In the PLS-PM literature each part of the estimated model needs to be validated: the measurement model, the structural model and the general model. Therefore, before analyzing the coefficients around human capital, the whole model must be validated to examine the degree to which reliable results are produced.

**Measurement model**

As the measurement models in this study are reflective, a unidimensionality analysis must be done, given the manifest variables are assumed to be caused by the same latent variable. For this, the Cronbach's alpha[1], Dillon-Goldstein's Rho[2] and eigenvalues[3] indexes are used, whose values are shown in table D1. Using the rule of values greater than 0.7 for Cronbach's alpha and Dillon-Goldstein's Rho indexes, it can be seen that the socioeconomic block does not reach the acceptable level for the first index. However, as pointed out by Chin (1998), the Dillon-Goldstein´s Rho is considered to be a better indicator than that of Cronbach's Alpha since the former is based on model results (i.e. loadings), rather than the correlations observed between the variables manifested in the database. Thus, by passing the tests on the next two indicators the blocks can be considered as unidimensionality.

**Table D1: Assessing Unidimensionality**

|  | Mode | MV | Cronbach | Dillon-Goldstein | First Eigenvalues | Second Eigenvalues |
|---|---|---|---|---|---|---|
| Socio-economic | A | 2 | 0.598 | 0.833 | 1.43 | 0.57 |
| Household size | A | 1 | 1.000 | 1.000 | 1.00 | 0.00 |
| Health status | A | 2 | 0.985 | 0.993 | 1.97 | 0.03 |
| Educ. achievements | A | 2 | 0.740 | 0.885 | 1.59 | 0.41 |
| Human capital | A | 2 | 0.908 | 0.956 | 1.83 | 0.17 |

Source: Author's preparation based on indicators from the World Bank and Barro and Lee (2013).
Note: Estimates are based on a reflective model using the year 1970 as reference for a sample of 91 countries.

---

[1] Cronbach's Alpha is a coefficient that measures how well a block of MV measures its corresponding LV. In this sense, each MV is assumed to be measured by the same LV and, therefore, these must be highly correlated between them. This coefficient is between 0 and 1, considering that values higher than 0.7 are sufficient to guarantee the unidimensionality of the block.

[2] It is another coefficient used to evaluate unidimensionality of a reflective block. This indicator focuses on the variance of the sum of the variables of the block of interest. As a general rule, a block is considered to be unidimensionality when this indicator is greater than or equal to 0.7.

[3] The third coefficient involves an analysis of eigenvalues of the correlation matrix of the set of indicators connected to a LV. If a block is unidimensional the first eigenvalue should be greater than 1, while in the second it should be less than 1.

However, in order to make the blocks more valid it is necessary to evaluate if the indicators are well explained by their latent variables. One way to do this is by examining their loadings or simple correlations of the indicators with their respective latent variable[4]. The most accepted empirical rule is the one proposed by Carmines and Zeller (1979), pointing out that the loading of an indicator is accepted as a member of a latent variable if it has a value equal to or greater than 0.7, which implies that more than 50 of the variance of an observed variable is shared by the construct. However, several researchers believe that this empirical rule ($\lambda \geq 0.7$) should not be so rigid in the initial stages of scale development (Barclay et al., 1995; Chin, 1998). In this sense Gefen et al. (2000) suggests that only the manifest variables with loadings greater than 0.4 are significant. Table D2 confirms that the blocks are well explained by the latent variables even in view of the requirements of Carmines and Zeller (1979).

Additionally, the communality index can be used. This index measures how much of the variability of the manifest variables in the *q*-th block is explained by its own latent variable. The ideal scenario is to have more shared variance between latent variables and their indicators than noise. Generally, "good" communality values are greater than 0.5 (Tenenhaus et al., 2005). Table D2 shows that the manifest variables of all blocks are well explained by their latent variables.

**Table D2: Evaluation of the validity of the results**

|   | VM | Block | Weights | Loadings | Communality | Redundancy |
|---|----|-------|---------|----------|-------------|------------|
| 1 | VAAS | Socio-economic | 0.549 | 0.819 | 0.699 | - |
| 2 | GNI | Socio-economic | 0.634 | 0.868 | 0.769 | - |
| 3 | FR | Household size | 1.000 | 1.000 | 1.000 | 0.506 |
| 4 | LE | Health status | 0.512 | 0.993 | 0.986 | 0.693 |
| 5 | MR | Health status | 0.496 | 0.992 | 0.985 | 0.693 |
| 6 | AYE | Educ. achievements | 0.663 | 0.930 | 0.865 | 0.651 |
| 7 | SPR | Educ. achievements | 0.454 | 0.844 | 0.712 | 0.536 |
| 8 | EC | Human capital | 0.523 | 0.957 | 0.916 | 0.694 |
| 9 | PP | Human capital | 0.522 | 0.957 | 0.915 | 0.693 |

Source: Author's preparation based on indicators from the World Bank and Barro and Lee (2013).
Note: Estimates are based on a reflective model using the year 1970 as reference for a sample of 91 countries.

Finally, table D3 shows the extent to which a particular construct differs from another, this is known as discriminant validity assessment. This is done by verifying that the shared variance between a block and its indicators is larger than the variance shared with other blocks

---

[4] The loading is the coefficient of regression that accompanies the LV in explaining the *p*-th MV in the *q*-th block. They represent the correlations between each MV and the corresponding LV. This evaluation is known as individual item reliability, those indicators that do not meet this criterion could be eliminated from the model.

(Sanchez, 2009). As can be seen, no indicator load higher on another construct than on the construct it is trying to measure.

**Table. D3: Discriminant Validity Assessment**

|   | Name | Block | Socio-economic | Household size | Health | Edu. | HC |
|---|------|-------|----------------|----------------|--------|------|-----|
| 1 | VAAS | Socio-economic | 0.819 | 0.466 | 0.674 | 0.593 | 0.529 |
| 2 | GNI | Socio-economic | 0.868 | 0.719 | 0.655 | 0.629 | 0.742 |
| 3 | FR | Household size | 0.712 | 1.000 | 0.766 | 0.768 | 0.794 |
| 4 | LE | Health status | 0.783 | 0.775 | 0.993 | 0.856 | 0.845 |
| 5 | MR | Health status | 0.776 | 0.745 | 0.992 | 0.822 | 0.817 |
| 6 | AYE | Educ. achievements | 0.760 | 0.780 | 0.870 | 0.930 | 0.846 |
| 7 | SPR | Educ. achievements | 0.485 | 0.552 | 0.592 | 0.844 | 0.603 |
| 8 | EC | Human capital | 0.751 | 0.735 | 0.797 | 0.805 | 0.957 |
| 9 | PP | Human capital | 0.714 | 0.784 | 0.806 | 0.793 | 0.957 |

Source: Author's preparation based on indicators from the World Bank and Barro and Lee (2013).
Note: Estimates are based on a reflective model using the year 1970 as reference for a sample of 91 countries.

**Structural model**

Now that the external model has been validated, the structural model can be examined. First, the coefficient of determination of endogenous latent variables $R^2$ is analyzed. Falk and Miller (1992) point out that the explained variance of endogenous variables should be greater than or equal to 0.1. These authors emphasize that values of $R^2$ less than 0.1, even though they are statistically significant, provide very little information, so the relationships that are formulated as hypotheses related to this latent variable have a very low predictive level. Table D4 shows that the latent variables in this study show a high predictive power. Additionally, as a general reference of the block, the average communality index can be used, which indicates the average amount of variance between a latent variable and its indicators that is common to both. Satisfactory latent variables should account for more than half of the variance, that is, the average communality index should exceed 0.5 (Tenenhaus et al., 2005). In the same line, the coefficient of determination is usually accompanied by an analogous measure, known as the redundancy index. As depicted in Vinzi et al. (2010), this index calculates the proportion of variability of the manifest variables in a given block, as a consequence of the latent variables connected to it. The test usually used to evaluate convergence validity in a block is known as Average Variance Extracted (AVE). This index tries to measure the amount of variance that a latent variable captures from its indicators in relation to the amount of variance due to the error measure. According to Fornell and Larcker (1981), this index should exceed 0.5. However, when the variables are standardized, the AVE measure is similar to that of average communality.

**Table D4: Inner Model Assessment**

|  | Type | $R^2$ | Av. Commu. | Av. Redun. | AVE |
|---|---|---|---|---|---|
| Socio-economic | Exogenous |  | 0.712 |  | 0.712 |
| Household size | Exogenous | 0.506 | 1.000 | 0.506 | 1.000 |
| Health status | Endogenous | 0.704 | 0.985 | 0.693 | 0.985 |
| Educ. achievements | Endogenous | 0.752 | 0.788 | 0.593 | 0.788 |
| Human capital | Endogenous | 0.758 | 0.916 | 0.694 | 0.916 |

Source: Author's preparation based on indicators from the World Bank and Barro and Lee (2013).
Note: Estimates are based on a reflective model using the year 1970 as reference for a sample of 91 countries.

Although the quality of each structural model can be measured by a simple evaluation of the coefficient of determination $R^2$, this is not sufficient to evaluate the whole structural model. In particular, since the structural equations are estimated once the convergence is achieved and the values of the latent variables are estimated, then $R^2$ takes into account only the fit of each regression equation in the structural model. A criterion of global goodness of fit has been developed with the objective of taking into account the behavior of the model, both the measurement and the structural model, this index is known as the *GoF*[5]. In the literature values greater than 0.7 are considered good within the literature of PLS-PM. The model presented in this study shows a *GoF* = 0.7604.

---

[5] This indicator is a nonparametric test that helps to evaluate the behavior of both the internal and the external model, i.e. it helps to evaluate the predictive behavior of the global model. The GoF is calculated as the geometric mean of the average communality and the average value of $R^2$ (Vinzi, Trinchera, and Amato, 2010).

## Sensitivity analysis

As defined by equation 1 to 4, health was allowed to influence education. This causal relationship can also be reversed (see, for example, Cawley and Ruhm, 2011; Cutler, 2006). Tables E1 and E2 show sensitivity analyses of the models presented in tables 1 and 2 that allow education to affect health.

**Table E1: Base structural model: Sensitivity analysis**

|   | Coefficient | Direct | Indirect | Total |
|---|---|---|---|---|
| Socio-economic → Household size | -0.712*** | -0.712 |  | -0.712† |
|  | (0.0745) |  |  | (0.045) |
| Coefficient of determination, $R^2$ | 0.506 |  |  |  |
| Household size → Educ. Achievements | -0.511*** |  |  | -0.511† |
|  | (0.089) |  |  | (0.078) |
| Socio-economic → Educ. Achievements | 0.360*** | 0.360 | 0.364 | 0.724† |
|  | (0.089) |  |  | (0.041) |
| Coefficient of determination, $R^2$ | 0.653 |  |  |  |
| Household size → Health status | -0.166** | -0.166 | -0.253 | -0.419† |
|  | (0.082) |  |  | (0.094) |
| Socio-economic → Health status | 0.309*** | 0.079 | 0.476 | 0.785† |
|  | (0.076) |  |  | (0.033) |
| Educ. achievements → Health status | 0.494*** | 0.494 |  | 0.494† |
|  | (0.084) |  |  | (0.092) |
| Coefficient of determination, $R^2$ | 0.788 |  |  |  |
| Household size → Human capital |  |  | -0.421 | -0.421† |
|  |  |  |  | (0.072) |
| Socio-Economic → Human capital |  |  | 0.684 | 0.684† |
|  |  |  |  | (0.041) |
| Health status → Human capital | 0.461*** | 0.461 |  | 0.461† |
|  | (0.098) |  |  | (0.083) |
| Educ. achievements → Human capital | 0.445*** | 0.445 | 0.228 | 0.673† |
|  | (0.098) |  |  | (0.053) |
| Coefficient of determination, $R^2$ | 0.758 |  |  |  |
| *GoF* | 0.7584 | 0.7577 |  |  |

Source: Author's preparation based on indicators from the World Bank and Barro and Lee (2013).
Note: Model is estimated for standardized latent variables. Standard errors are in parentheses. **$p < 0.05$, ***$p < 0.01$.
†Validation of the significance of the coefficients of the total effects is based on 95% confidence intervals using the bootstrap resampling method.

**Table E2: Modified structural model: Sensitivity analysis**

|  | Coefficient | Direct | Indirect | Total |
|---|---|---|---|---|
| Socio-economic → Household size | -0.709*** | -0.709 |  | -0.709† |
|  | (0.075) |  |  | (0.046) |
| Coefficient of determination, $R^2$ | 0.503 |  |  |  |
| Household size → Educ. Achievements | -0.511*** | -0.511 |  | -0.511† |
|  | (0.089) |  |  | (0.080) |
| Socio-economic → Educ. Achievements | 0.361*** | 0.361 | 0.363 | 0.724† |
|  | (0.089) |  |  | (0.040) |
| Coefficient of determination, $R^2$ | 0.654 |  |  |  |
| Household size → Health status | -0.167** | -0.167 | -0.252 | -0.419† |
|  | (0.082) |  |  | (0.085) |
| Socio-economic → Health status | 0.311*** | 0.311 | 0.475 | 0.786† |
|  | (0.076) |  |  | (0.034) |
| Educ. achievements → Health status | 0.492*** | 0.492 |  | 0.492† |
|  | (0.084) |  |  | (0.089) |
| Coefficient of determination, $R^2$ | 0.789 |  |  |  |
| Household size → Human capital | -0.168** | -0.168 | -0.187 | -0.355† |
|  | (0.069) |  |  | (0.071) |
| Socio-Economic → Human capital | 0.465*** | 0.465 | 0.405 | 0.871† |
|  | (0.068) |  |  | (0.024) |
| Health status → Human capital | 0.109 | 0.109 |  | 0.109 |
|  | (0.088) |  |  | (0.098) |
| Educ. achievements → Human capital | 0.277*** | 0.277 | 0.054 | 0.331† |
|  | (0.081) |  |  | (0.069) |
| Coefficient of determination, $R^2$ | 0.861 |  |  |  |
| $GoF$ | 0.7584 |  |  |  |

Source: Author's preparation based on indicators from the World Bank and Barro and Lee (2013).
Note: Model is estimated for standardized latent variables. Standard errors are in parentheses. **$p< 0.05$, ***$p< 0.01$. †Validation of the significance of the coefficients of the total effects is based on 95% confidence intervals using the bootstrap resampling method.

The results support this dual causality; therefore, education has a strong and significant impact on health. However, the overall results of the general model change marginally, showing that the model remains robust regardless of the causality established between these two variables. Regarding total effects, the impact of education increases because it gains the indirect effects generated on health, approaching from the socio-economic context. Table E2 shows that the effect is less important when all variables simultaneously influence the return on human capital. In this case, education is equated to the effects of household resources and, to a lesser extent, to the effects of socio-economic contexts.

On the other hand, as mentioned above, the estimated index could be biased because when one extracts the variance corresponding to human capital using manifest variables for EC (as a proxy for productivity) and PP (as a proxy for the innovative and inventive capacity of individuals), one might consider only a portion of the distribution. This problem should be solved if the block includes manifest variables that are also highly correlated with human capital but incorporate elements that are not present in the variables used thus far. The problem with this approach is that there are no data at the international level for the period of time and number of countries included in the present study. However, to assess the degree of bias of the model, the results for a single period are analyzed with the addition of two new variables to the HCI block.

First, the variable trade per capita of equipment related to research and education is used. This variable recognizes the importance of information technology and the necessity of information for the application of cognitive and research abilities[6]. Although this variable is available for all countries referenced, it is only available for the period 1970-2000.

Second, an international test score variable is included to account for students' learning achievements. The literature has justified the use of this variable in human capital estimations because these tests are closely linked to innovation and productivity, both theoretically and empirically (Hanushek and Kimko, 2000). In this document, the data from Altinok and Murseli (2007) are used. The authors built a cognitive indicator with a sample of various international tests. The problem with this variable is that data are only consistently available for the years 2000, 2003, 2005, 2007 and 2009. In addition, this indicator includes only a small number of developing countries (22 countries, of which two are African and six are Latin American), which reduces the sample to 44 countries. Given the limited data availability, the sensitivity analysis is performed on a year for which data for the four variables are available, namely, the year 2000. Table E3 shows different specifications that include the two variables mentioned above, compared with the model as defined by equation 1 to 4. Table E3 presents the weights, loadings, and coefficients of determination $R^2$ for the HCI block, as well as the goodness of fit of the model and the size of the sample.

Model 1, which was presented by equations 1 to 4, is provided for this common year. The variable trade per capita of computer equipment and information related to research and education (T-R&D) is added in model 2. The inclusion of this variable reduces the weights of EC and PP by 0.159 and 0.153, respectively, which influences the HCI scores (see table E4). However, the loadings of these variables change only marginally (0.004 and 0.057, respectively) and maintain a significant association with the block (Cronbach's alpha=0.918 and Dillon-Goldstein's rho=0.948). In fact, the results of both models show a high coefficient of association

---

[6] This variable is the sum of imports and exports of information and communications technology (ICT) equipment in $US. The data are from Feenstra et al. (2005) for the 7511-7529 classes of the SITC codes Rev. 2 (4 digits). For a justification of the use of this variable in relation to human capital, see Apergis et al. (2009) and Hanushek and Wößmann (2008).

($\rho$ =0.91). However, his new variable does not seem to have much influence in terms of explaining the block or on the goodness of fit of the model.

**Table E3: Sensitivity analysis for HCI with inclusion of C-R&D and ICA**

|  | Model 1 | | Model 2 | | | Model 3 | |
| --- | --- | --- | --- | --- | --- | --- | --- |
|  | EC | PP | EC | PP | C-R&D | EC | PP |
| Weights | 0.542 | 0.499 | 0.383 | 0.346 | 0.348 | 0.534 | 0.529 |
| Loadings | 0.963 | 0.956 | 0.959 | 0.899 | 0.922 | 0.942 | 0.940 |
| $R^2$ | 0.775 | | 0.791 | | | 0.653 | |
| GoF | 0.7973 | | 0.7930 | | | 0.6588 | |
| n | 91 | | 91 | | | 44 | |

|  | Model 4 | | | Model 5 | | |
| --- | --- | --- | --- | --- | --- | --- |
|  | EC | PP | C-R&D | EC | PP | ICA |
| Weights | 0.394 | 0.390 | 0.333 | 0.372 | 0.368 | 0.390 |
| Loadings | 0.915 | 0.913 | 0.852 | 0.865 | 0.936 | 0.856 |
| $R^2$ | 0.642 | | | 0.742 | | |
| GoF | 0.6482 | | | 0.6618 | | |
| n | 44 | | | 44 | | |

|  | Model 6 | | | |
| --- | --- | --- | --- | --- |
|  | EC | PP | C-R&D | ICA |
| Weights | 0.298 | 0.292 | 0.252 | 0.312 |
| Loadings | 0.856 | 0.911 | 0.848 | 0.849 |
| $R^2$ | 0.717 | | | |
| GoF | 0.6517 | | | |
| n | 44 | | | |

Source: Author's preparation based on indicators from the World Bank, Barro and Lee (2013), Feenstra et al. (2005) and Altinok and Murseli (2007).
Note: Estimates are based on a reflective model using the year 2000 as reference for a sample of 91 countries (44 countries when the analysis includes cognitive test scores).

For purposes of making comparisons when the International Cognitive Assessment variable (ICA) is added, the sample is reduced to 44 countries in models 3 through 6. Model 3 is the base model for this country sample. The reduction of the sample strongly influences the coefficient of determination and the fit of the model but does not significantly affect the weights, loadings and scores. The incorporation of T-R&D into the model for the sample of 44 countries (model 4) produces results very similar to model 2; that is, the weights decrease but there are no significant changes to the rest of the indicators in model 3.

The incorporation of ICA into the model (model 5) causes several changes to model 3. First, the explanatory power of the model increases, particularly that of the block. The ICA variable, along with T-R&D, seems to identify the block more precisely by eliminating the noise present in the original model. ICA reduces the weights of EC and PP by 0.162 and 0.139, respectively, which can be attributed to the bias present in models 1 and 3. Nonetheless, the variables once again maintain relatively stable loadings, indicating a high association with the block, and thus correctly measure the returns on human capital (Cronbach's alpha=0.863 and Dillon-Goldstein's rho=0.917). Finally, model 6 incorporates the ICA variable into model 4. The incorporation of both of these variables reduces the weights of EC and PP by 0.236 and 0.237, respectively, but only marginally reduces the loadings. The results are similar to those of model 5 in terms of fit. In model 6, the scores of countries increase without significantly altering their relative positions (see table E5).

In summary, the incorporation of the ICA and T-R&D variables shows that the estimated indicator includes some bias that increases the weights of the manifest variables of the HCI block (on average, 0.18 for both variables). Note, however, that neither the loadings nor the explanatory power of the models are altered significantly. In fact, the inclusion of these two new variables shows that EC and PP are highly related to the block they attempt to measure and therefore explain the behavior of the block well. In addition, the models have high correlations, $\rho$ =0.98 for models 3 and 5 and $\rho$ =0.95 for models 3 and 6. These results indicate the validity of the estimated indicator, with the caveat of the present bias, which could be reduced with the use of manifest variables such as ICA, although the inclusion of the ICA variable restricts its cross-sectional use to a limited sample of countries.

**Table E4: Scores for the estimated HCI by country, $n$ = 91**

| | Upper HCI | | | | | | Lower HCI | | | | |
|---|---|---|---|---|---|---|---|---|---|---|---|
| | Model 1 | | | Model 2 | | | Model 1 | | | Model 2 | |
| Rank | Country | Score | Rank | Country | Score | Rank | Country | Score | Rank | Country | Score |
| 1 | Japan | 100.0 | 1 | Japan | 100.0 | 82 | Sudan | 0.095 | 82 | Sudan | 0.068 |
| 2 | Korea | 62.42 | 8 | Korea | 65.7 | 83 | Nigeria | 0.078 | 81 | Nigeria | 0.081 |
| 3 | United States | 59.84 | 7 | United States | 68.7 | 84 | Nepal | 0.076 | 80 | Nepal | 0.081 |
| 4 | Finland | 58.85 | 4 | Finland | 74.4 | 85 | Cameroon | 0.076 | 87 | Cameroon | 0.038 |
| 5 | Norway | 58.78 | 6 | Norway | 72.0 | 86 | Senegal | 0.058 | 86 | Senegal | 0.039 |
| 6 | Sweden | 58.34 | 5 | Sweden | 72.4 | 87 | Togo | 0.056 | 90 | Togo | 0.030 |
| 7 | Iceland | 47.95 | 14 | Iceland | 53.0 | 88 | Côte d'Ivoire | 0.056 | 88 | Côte d'Ivoire | 0.038 |
| 8 | Germany | 42.57 | 11 | Germany | 57.4 | 89 | Tanzania | 0.053 | 84 | Tanzania | 0.048 |
| 9 | New Zealand | 40.02 | 17 | New Zealand | 42.1 | 90 | Benin | 0.043 | 91 | Benin | 0.026 |
| 10 | Luxembourg | 38.01 | 10 | Luxembourg | 59.0 | 91 | Ethiopia | 0.036 | 89 | Ethiopia | 0.030 |

Source: Author's preparation based on indicators from the World Bank, Barro and Lee (2013), Feenstra et al. (2005) and Altinok and Murseli (2007).
Note: Manifest variables were adjusted to the 0-100 scale, with 100 as the highest possible value. For reasons of space, only the top and bottom of the estimates of the HCI scores are shown. Estimates are based on a reflective model using the year 2000 as reference for a sample of 91 countries (44 countries when the analysis includes cognitive tests).

**Table E5: Scores for the estimated HCI by country, *n* = 44**

| Upper HCI | | | | | | | | | | | |
|---|---|---|---|---|---|---|---|---|---|---|---|
| **Model 3** | | | **Model 4** | | | **Model 5** | | | **Model 6** | | |
| Rank | Country | Score | Rank | Country | Score | Rank | Country | Score | Rank | Country | Score |
| 1 | Japan | 100 | 1 | Japan | 100.0 | 1 | Japan | 100 | 1 | Japan | 100 |
| 2 | Korea | 62.0 | 8 | Korea | 65.0 | 2 | Korea | 73.6 | 6 | Korea | 73.5 |
| 3 | USA | 57.6 | 7 | USA | 65.7 | 3 | USA | 66.5 | 8 | USA | 71.3 |
| 4 | Finland | 56.3 | 4 | Finland | 70.3 | 4 | Finland | 66.3 | 4 | Finland | 75.6 |
| 5 | Sweden | 55.8 | 5 | Sweden | 68.5 | 5 | Sweden | 65.2 | 5 | Sweden | 73.6 |
| 6 | Norway | 55.2 | 6 | Norway | 67.2 | 6 | Norway | 63.5 | 7 | Norway | 71.5 |
| 7 | Iceland | 44.7 | 12 | Iceland | 49.4 | 7 | Iceland | 57.5 | 12 | Iceland | 59.0 |
| 8 | New Zealand | 38.5 | 15 | New Zealand | 40.4 | 8 | New Zealand | 50.7 | 15 | New Zealand | 49.7 |
| 9 | Luxembourg | 35.8 | 10 | Luxembourg | 54.1 | 9 | Luxembourg | 47.9 | 11 | Luxembourg | 61.3 |
| 10 | Canada | 31.2 | 14 | Canada | 43.6 | 11 | Canada | 45.3 | 14 | Canada | 53.8 |
| Lower HCI | | | | | | | | | | | |
| **Model 3** | | | **Model 4** | | | **Model 5** | | | **Model 6** | | |
| Rank | Country | Score | Rank | Country | Score | Rank | Country | Score | Rank | Country | Score |
| 35 | Jordan | 2.92 | 38 | Jordan | 2.29 | 35 | Jordan | 8.9 | 38 | Jordan | 6.0 |
| 36 | Thailand | 2.34 | 30 | Thailand | 5.43 | 36 | Thailand | 8.0 | 28 | Thailand | 11.7 |
| 37 | Iran | 2.03 | 41 | Iran | 1.28 | 37 | Iran | 7.0 | 41 | Iran | 4.0 |
| 38 | México | 1.77 | 35 | México | 3.86 | 38 | México | 6.5 | 35 | México | 9.2 |
| 39 | Turkey | 1.76 | 37 | Turkey | 2.30 | 39 | Turkey | 6.4 | 37 | Turkey | 6.3 |
| 40 | Tunisia | 1.45 | 39 | Tunisia | 1.85 | 40 | Tunisia | 5.7 | 39 | Tunisia | 5.3 |
| 41 | Colombia | 0.83 | 42 | Colombia | 1.23 | 41 | Colombia | 3.7 | 42 | Colombia | 4.0 |
| 42 | Peru | 0.68 | 43 | Peru | 1.01 | 42 | Peru | 3.2 | 43 | Peru | 3.5 |
| 43 | Philippines | 0.66 | 40 | Philippines | 1.77 | 43 | Philippines | 3.2 | 40 | Philippines | 5.2 |
| 44 | Indonesia | 0.37 | 44 | Indonesia | 0.61 | 44 | Indonesia | 2.3 | 44 | Indonesia | 2.8 |

Source: Author's preparation based on indicators from the World Bank, Barro and Lee (2013), Feenstra et al. (2005) and Altinok and Murseli (2007).
Note: The manifest variables were adjusted to the 0-100 scale, with 100 as the highest possible value. For reasons of space, only the top and bottom of the estimates of the HCI scores are shown. Estimates are based on a reflective model using the year 2000 as reference for a sample of 91 countries (44 countries when the analysis includes cognitive tests).

**Testing the Human Capital Index**

In this section a test between HCI and those standard variables used for measuring human capital is performed. A first look at the performance of the indicator is undertaken by associating it with the traditional measures for human capital (See table F1). The correlation between AYE and the proposed index (*HCI*) is 0.86 for a sample of 91 countries and 0.72 when the sample is reduced to 44. The sample is reduced to 44 countries in order to homogenize it when it is compared with ICA, which is consistently available for a smaller number of countries and periods.

The problem with ICA is that it is available to a small number of countries and periods. Taking data from Altinok and Murseli (2007) and the year 2000 as a reference (a year in which all variables are available), it is shown that *HCI* and ICA have a correlation coefficient of 0.83, which is higher than the relation between the two educational variables, which is 0.60. On the other hand, note that ICA could be a *HCI* block element, as this is part of the returns on human capital. In *HCI2* the proposed indicator is re-estimated to include in the block of returns on human capital to ICA as a manifest variable. As expected, the relation between the previous measure (*HCI*) and *HCI2* have a strong association (0.92) in spite of the sample in this process being reduced to 44 countries.

**Table No. F1: correlations between different measures of human capital**

|      | *hc*    | *hc2*   | *AYE*   | *ICA*   |
|------|---------|---------|---------|---------|
| *AYE* | 0.857$^a$ | 0.793$^a$ |         |         |
| *ICA* |         | 0.735$^a$ | 0.603$^a$ |         |
| *hc*  |         | 0.921$^a$ | 0.721$^a$ | 0.829$^a$ |
|      | $n = 91$ | $n = 44$ | $n = 44$ | $n = 44$ |

Note: $^a$ Spearman´s rho with significance level of 1%. *HC* is the proposed indicator inserted in the block of human capital returns with manifest variables: patents per capita and energy consumption per capita. *HCI2* is the same index plus one variable of trade per capita related research and education equipment (The data are from Feenstra et al., 2005) and an international cognitive assessment variable (ICA) from Altinok and Murseli (2007).

On the other hand, given the caveats mentioned in the main document and based on the econometric specification of Mankiw, Romer, and Weil (1990) in which GDP per worker is used as the dependent variable and different regressors as its determinants, table F2 shows the results for the different variables of human capital in explaining growth. Columns (1) through (4) are regressions of each of the variables without the use of controls; in subsequent columns investment in physical capital and population growth rate, among others[7], are used as main controls.

---

[7] The variables included are: the investment in physical capital, measured as the average share of investment real to GDP, population growth rate, average government consumption as a percentage of GDP, variables taken from PWT; inflation measured by consumer prices from the World Bank indicators; A binary variable measuring the level of democracy in the countries and two estimated indicators by principal component analysis to approximate the degree of impugnment of the countries, these three last variables which come from the Teorell et al. (2013).

As noted, all measures of human capital show the expected sign and are highly significant. However, the proposed variables exhibit greater explanatory power, both in models in which no controls are included, as when they are present. For instance, under control, *HCI* and *HCI2* increase $R^2$ by 16 percentage points relative to ICA and 17 points with AYE. Despite the proper functioning of the proposed variables, these results are only an initial approximation to the performance of *HCI* explaining economic growth and therefore these results should be viewed with caution, given the diverse problems present in the estimates in Table F2. Further analysis should be performed in order to eliminate most of the present bias in these estimates.

**Table No. F2: Different human capital measures for basic regression analysis as determinants of economic growth by OLS**

| | Dependent variable: GDP per worker | | | | | | | |
|---|---|---|---|---|---|---|---|---|
| | (1) | (2) | (3) | (4) | (5) | (6) | (7) | (8) |
| ln(AYE) | 1.912*** | | | | 1.091** | | | |
| | (0.424) | | | | (0.477) | | | |
| ln(ICA) | | 10.115*** | | | | 7.097** | | |
| | | (2.616) | | | | (3.256) | | |
| ln(HCI) | | | 0.389*** | | | | 0.373*** | |
| | | | (0.042) | | | | (0.060) | |
| ln(HCI2) | | | | 3.528*** | | | | 3.727*** |
| | | | | (0.3915) | | | | (0.572) |
| Controls | No | No | No | No | Si | Si | Si | Si |
| Sample size | 44 | 44 | 44 | 44 | 44 | 44 | 44 | 44 |
| $R^2$ | 0,341 | 0,279 | 0,596 | 0.629 | 0.488 | 0.500 | 0.663 | 0.677 |

Note: Robust standard errors in parentheses. ** $p<0.05$, *** $p<0.01$. Columns 1 to 4 show comparisons between the two proposed indicators (*HCI* and *HCI2*) and two educational variables (AYE and ICA) when no controls are included. Columns 5 to 8 present the same comparisons when the regressions include controls. Control variables: investment in physical capital, population growth rate, average government consumption, inflation, a binary variable measuring the level of democracy in the countries and two estimated indicators by principal component analysis to approximate the degree of impugnment of the countries.